\documentclass[structabstract]{aa}  

\usepackage{graphicx}
\usepackage{url}
\usepackage{txfonts}
\usepackage{multirow}
\usepackage{aalongtable}
\usepackage[]{natbib}
\bibpunct{(}{)}{;}{a}{}{,}

\begin{document}

   \title{GRB 091029: At the limit of the fireball scenario}

   \author{R. Filgas\inst{1,2} 
           \and J. Greiner\inst{1} 
           \and P. Schady\inst{1}
           \and A. de Ugarte Postigo\inst{3,4}
           \and S. R. Oates\inst{5}
           \and M. Nardini\inst{6,1}
           \and T. Kr{\"u}hler\inst{4,1,7}
           \and A. Panaitescu\inst{8}
           \and D. A. Kann\inst{9,1,7}
           \and S. Klose\inst{9}
           \and P. M. J. Afonso\inst{1}\thanks{Present address: American River College, Physics \& Astronomy Dpt., 4700 College Oak Drive, Sacramento, CA 95841}
           \and W. H. Allen\inst{10} 
           \and A. J. Castro-Tirado\inst{3}
           \and G. W. Christie\inst{11}
           \and S. Dong\inst{12}
		   \and J. Elliott\inst{1}
           \and T. Natusch\inst{13} 
           \and A. Nicuesa Guelbenzu\inst{9}           
           \and F. Olivares E.\inst{1} 
           \and A. Rau\inst{1}  
           \and A. Rossi\inst{9} 
           \and V. Sudilovsky\inst{1}
           \and P.C.M. Yock\inst{14}
            }

   \institute{Max-Planck-Institut f\"ur extraterrestrische Physik, Giessenbachstra\ss{}e 1, 85748 Garching, Germany, \\     
     \email{filgas@mpe.mpg.de}
     \and Institute of Experimental and Applied Physics, Czech Technical University in Prague, Horsk\'a 3a/22, 12800 Prague, Czech Republic
     \and Instituto de Astrof\'isica de Andaluc\'ia (IAA-CSIC), Glorieta de la Astronom\'ia s/n, 18008 Granada, Spain      
     \and Dark Cosmology Centre, Niels Bohr Institute, University of Copenhagen, Juliane Maries Vej 30, 2100 Copenhagen, Denmark  
     \and Mullard Space Science Laboratory, University College London, Holmbury St. Mary, Dorking Surrey, RH5 6NT, UK    
     \and Universit\`a degli studi di Milano-Bicocca, Piazza della Scienza 3, 20126, Milano, Italy
     \and Universe Cluster, Technische Universit\"at M\"unchen, Boltzmannstra\ss{}e 2, 85748 Garching, Germany
     \and Space Science and Applications, MS D466, Los Alamos National Laboratory, Los Alamos, NM 87545, USA
     \and Th\"uringer Landessternwarte Tautenburg, Sternwarte 5, 07778 Tautenburg, Germany
     \and Vintage Lane Observatory, Blenheim, New Zealand
     \and Auckland Observatory, P.O. Box 12-180, Auckland, New Zealand
     \and Institute for Advanced Study, Einstein Drive, Princeton, NJ 08540, USA
     \and AUT University, Auckland, New Zealand
     \and Department of Physics, University of Auckland, Auckland, New Zealand
        }
      
   \date{Received 11 May 2012 / Accepted 13 September 2012}

  \abstract
   {}
   {Using high-quality, broad-band afterglow data for GRB~091029, we 
    test the validity of the forward-shock model for gamma-ray burst afterglows.}
   {We used multi-wavelength (NIR to X-ray) follow-up observations obtained
    with the GROND, BOOTES-3/YA and Stardome optical ground-based telescopes, 
    and the UVOT and the XRT onboard the \emph{Swift} satellite. 
    The resulting data of excellent accuracy allow us to construct a multi-wavelength light curve 
    with relative photometric errors as low as 1\%, as well as 
    the well-sampled spectral energy distribution covering 5 decades in energy.}
   {The optical/NIR and the X-ray light curves of the afterglow of GRB 091029 are almost totally decoupled. 
    The X-ray light curve shows a shallow rise with a peak at $\sim7$~ks and a decay slope of $\alpha\sim1.2$ afterwards, 
    while the optical/NIR light curve shows a much steeper early rise with a peak around $400$~s, followed by a shallow decay with 
    temporal index of $\alpha\sim 0.6$, a bump and a steepening of the decay afterwards. 
    The optical/NIR spectral index decreases gradually by over $0.3$ before this bump, and then slowly 
    increases again, while the X-ray spectral index remains constant throughout the observations.}
   {To explain the decoupled light curves in the X-ray and optical/NIR domains, a two-component outflow is proposed. Several models are tested, 
    including continuous energy injection, components with different electron energy indices and components in two different stages of
    spectral evolution. Only the last model can explain both the decoupled light curves with asynchronous peaks and the peculiar SED evolution.
    However, this model has so many unknown free parameters that we are unable to reliably confirm or disprove its validity, making the afterglow
    of GRB 091029 difficult to explain in the framework of the simplest fireball model.  
    This conclusion provides evidence that a scenario beyond the simplistic assumptions is needed to be
    able to model the growing number of well-sampled afterglow light curves.}
   \maketitle 
   
\section{Introduction}
   Since the first Gamma-Ray Burst (GRB) was discovered in the late 1960's \citep{1973ApJ...182L..85K}, the GRB field has evolved rapidly,
   mainly thanks to three generations of dedicated satellites. The \emph{Compton Gamma-Ray Observatory} was launched in 1991
   and with its instrument BATSE showed that GRBs are isotropically distributed in the sky, suggesting they might
   have a cosmological origin \citep{1992Natur.355..143M}. This claim was confirmed in 1997 by \emph{Beppo-SAX}, an 
   Italian-Dutch satellite that detected and precisely localized GRBs at X-ray wavelengths \citep{1997Natur.387..783C} and 
   enabled ground-based telescopes to perform follow-up observations \citep{1997Natur.386..686V}, including redshift measurements 
   that confirmed the cosmological distances of these events \citep{1997Natur.387..878M,1998Natur.393...35K}. 
   Finally, the \emph{Swift} satellite \citep{2004ApJ...611.1005G}, launched in 2004, allows for a study of the afterglow emission 
   starting very early after the GRB is detected by the Burst 
   Alert Telescope \citep[BAT,][]{2005SSRv..120..143B}, thanks to the rapid slewing capability of the spacecraft and a precise localization of the afterglow by 
   onboard telescopes sensitive at X-ray \citep[XRT,][]{2005SSRv..120..165B} and ultraviolet/optical \citep[UVOT,][]{2005SSRv..120...95R} wavelengths.
   Such precise and early localization allows ground-based follow-up telescopes to start observing the afterglow emission within tens of seconds of the burst
   onset. 
   
   The leading model for the afterglow emission is the fireball model \citep{1997ApJ...476..232M,1999PhR...314..575P,2002ARA&A..40..137M,
   2004IJMPA..19.2385Z}, where the afterglow arises from the synchrotron emission of shock-accelerated electrons in a  
   fireball interacting with the circum-burst medium. While most of the afterglow light curves prior to \emph{Swift} were consistent with this
   model \citep{1999ApJ...517L.105H,1999ApJ...522L..39S}, the more recent and detailed light curves of afterglows showed features that needed various
   additions and modifications to the simplest fireball model. The early steep decays of the optical light curves
   \citep{1999Natur.398..400A,2003Natur.422..284F,2003ApJ...586L...5F,2004ApJ...601.1013R} are interpreted as reverse shocks 
   \citep{1999ApJ...517L.109S,1999ApJ...520..641S,1999MNRAS.306L..39M,2000ApJ...545..807K}. Some of the rebrightenings and bumps
   \citep{1998ApJ...503..314P,2003Natur.426..138G,2007A&A...461...95G} are attributed to refreshed shocks \citep{1998ApJ...496L...1R,1998ApJ...503..314P}, 
   the others to density variations \citep{2001MNRAS.327..829R,2002ApJ...565L..87D,2003ApJ...591L..21D,2004MNRAS.353..511P} or two-component jets
   \citep{2003Natur.426..154B,2004ApJ...605..300H,2005ApJ...626..966P,2006MNRAS.370.1946G,2008Natur.455..183R,2011A&A...526A.113F}.
   However, with the latest generation of GRB instruments capable of high sampling in both time and energy domains, the 
   modifications made to the standard model still fall short to explaining the observed afterglows consistently \citep[e.g., ][]{2011A&A...531A..39N,2011A&A...535A..57F}.

   The Gamma-Ray burst Optical Near-infrared Detector \citep[GROND,][]{2008PASP..120..405G,2007Msngr.130...12G}
   has provided high-quality, very well-sampled, simultaneous data in seven bands since 2007, when it was
   mounted at the 2.2~m MPI/ESO telescope at La Silla observatory in Chile. 
   The high-precision data obtained by GROND allow for a detailed study of afterglow 
   light curves \citep{2009ApJ...693.1912G,2011A&A...531A..39N}, jets of GRBs \citep{2009A&A...508..593K},
   the dust in their host galaxies \citep{2008ApJ...685..376K,2010A&A...515L...2K,2011A&A...526A..30G,2011A&A...534A.108K,2012A&A...537A..15S}, 
   their redshifts \citep{2009ApJ...693.1610G,2011A&A...526A.153K}, their associations with SNe \citep{2012A&A...539A..76O}, and provide tools to test the 
   standard fireball scenario and its modifications.  

   Here we provide details of the \emph{Swift}/XRT, \emph{Swift}/UVOT, GROND, BOOTES-3 and Stardome observations of the 
   afterglow of GRB~091029 and discuss the light curves and spectral energy distributions (SEDs) in the context of the fireball shock model
   thanks to the very good energy and time-domain coverage of our high-quality data.    
   Throughout the paper, we adopt the convention that the flux density 
   of the GRB afterglow can be described as $F_\nu (t) \propto t^{-\alpha} \nu^{-\beta}$, where $\alpha$ is 
   the temporal and $\beta$ the spectral index. Unless stated otherwise in the text, all reported errors are at 1$\sigma$ confidence 
   level.
   
\section{Observations}
 \subsection{Swift}
   The \emph{Swift}/BAT was triggered by the long GRB~091029 at $T_0 =$ 03:53:22 UT and started 
   immediately slewing to the burst \citep{2009GCN..10097...1G}. The mask-weighted light curve 
   shows three overlapping peaks, starting at $~T_0-10$~s and ending at $~T_0+70$~s, with peaks
   at $T_0+2$, +20, and +26~s. The measured $T_{90}$ (15-350 keV) is $39.2 \pm 5$~s 
   \citep{2009GCN..10103...1B}. 
   The BAT prompt emission spectrum from $T_0-1.8$ to $T_0+60.2$~s is best fitted using a power-law
   with an exponential cutoff.  This fit gives a photon index of $1.46 \pm 0.27$ and 
   an $E_{\rm peak}=61.4 \pm 17.5$~keV.  For this model the total fluence in the 15-150 keV energy
   range is $2.4\pm0.1 \times 10^{-6}$~erg/cm$^2$ \citep{2009GCN..10103...1B}.
   Using standard concordance cosmology ($H_0 = 71.0$~km/s/Mpc, 
   $\Omega_M$ = 0.27, $\Omega_{\Lambda} = 0.73$, \citealt{2009ApJS..180..330K}), and a redshift 
   of $z=2.752$ \citep{2009GCN..10100...1C}, the bolometric (1keV - 10MeV) energy release of
   GRB~091029 is $E_{\rm iso} = 8.3 \times 10^{52}$~erg, with a rest-frame $E_{\rm peak}$ of $\sim$230~keV. 
   
   The \emph{Swift}/XRT started observations of the field of GRB~091029 79.3~s after the trigger 
   \citep{2009GCN..10097...1G}. XRT data were obtained from the public \textit{Swift} 
   archive and reduced in the standard manner using the xrtpipeline 
   task from the HEAsoft package, with response matrices from the most recent CALDB release. The XRT light 
   curve was obtained from the XRT light curve repository \citep{2007A&A...469..379E,2009MNRAS.397.1177E}.
   
   The \emph{Swift}/UVOT began settled observations of the field of GRB~091029 91~s after the trigger 
   \citep{2009GCN..10108...1M}. The afterglow was detected in the $White$, $U$, $B$ and $V$ filters. For this
   analysis, we have reduced both image and event mode data. Before the count rates were extracted from the
   event lists, the astrometry was refined following the methodology in \citet{2009MNRAS.395..490O}.
   The photometry was then extracted from the event lists and image files using the FTOOLs {\it uvotevtlc} and 
   {\it uvotmaghist}, respectively, using a source aperture centered on the optical position
   and a background region located in a source-free region.
   We used a $3^{\prime\prime}$ source aperture to avoid contamination from two neighbouring stars and applied aperture 
   corrections to the photometry in order to be compatible with the UVOT calibration \citep{2011AIPC.1358..373B}. 
   The analysis pipeline used software HEADAS 6.10 and UVOT calibration 20111031. In order to be compatible
   with the GROND photometry, UVOT magnitudes are provided as AB magnitudes and listed in Table \ref{091029uvot}.

 \begin{figure}[h]
   \resizebox{\hsize}{!}{\includegraphics[angle=270]{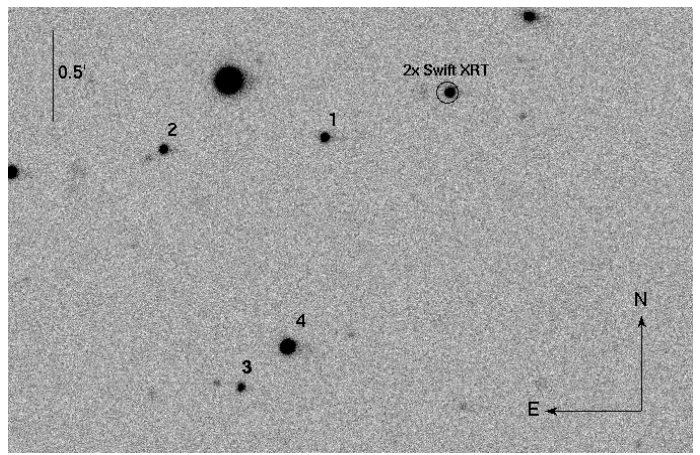}}
   \caption{GROND $g'$-band image of the field of GRB 091029 obtained  
             463~s after $T_0$. The optical afterglow is shown inside the \emph{Swift} XRT
             error circle with double diameter for better clarity. 
             The secondary standard stars are numbered from $1$ to $4$ and
             their magnitudes reported in Table \ref{standards091029}.}
   \label{chart091029}
 \end{figure} 
     
\begin{table*}
\caption{Secondary standards in the GRB 091029 field in the GROND
filter bands used for the calibration}             
\label{standards091029}      
\centering                         
\begin{tabular}{c c c c c c}        
\hline\hline                 
Star & R.A., Dec & $g'$ & $r'$ & $i'$ & $z'$ \\
 number & [J2000] & (mag$_\mathrm{AB}$) & (mag$_\mathrm{AB}$) & (mag$_\mathrm{AB}$) & (mag$_\mathrm{AB}$) \\
\hline                        
1 & 04:00:47.46, $-$55:57:35.1 & $18.50 \pm 0.04$ & $17.03 \pm 0.03$ & $16.31 \pm 0.06$ & $15.92 \pm 0.04$ \\      
2 & 04:00:53.70, $-$55:57:39.2 & $18.63 \pm 0.04$ & $18.14 \pm 0.03$ & $17.98 \pm 0.06$ & $17.85 \pm 0.05$ \\
3 & 04:00:50.68, $-$55:58:57.3 & $19.25 \pm 0.04$ & $18.95 \pm 0.03$ & $18.92 \pm 0.07$ & $18.89 \pm 0.06$ \\
4 & 04:00:48.90, $-$55:58:43.8 & $16.28 \pm 0.04$ & $15.75 \pm 0.03$ & $15.66 \pm 0.06$ & $15.55 \pm 0.04$ \\
\hline                                   
\end{tabular}
\\ [5pt]
\begin{tabular}{c c c c c}        
\hline\hline  
Star & R.A., Dec & $J$ & $H$ & $K_s$ \\
 number & [J2000] & (mag$_\mathrm{Vega}$) & (mag$_\mathrm{Vega}$) & (mag$_\mathrm{Vega}$) \\
\hline                     
1 & 04:00:37.34, $-$56:01:20.6 & $13.03\pm0.03$ & $12.67\pm0.03$ & $12.62\pm0.03$\\
2 & 04:00:39.43, $-$55:56:02.0 & $12.95\pm0.03$ & $12.65\pm0.03$ & $12.58\pm0.03$\\
3 & 04:00:45.75, $-$55:55:34.7 & $13.35\pm0.03$ & $13.07\pm0.03$ & $12.96\pm0.03$\\
4 & 04:00:47.49, $-$55:57:35.0 & $14.66\pm0.03$ & $13.98\pm0.03$ & $13.80\pm0.03$ \\
\hline                                   
\end{tabular}
\end{table*}  
 
 \subsection{GROND} 
   GROND responded to the \emph{Swift} GRB alert and initiated automated
   observations at 03:57~UT, 4.5~min after the trigger \citep{2009GCN..10098...1F},
   and imaged the field of GRB~091029 at seven later epochs until $T_0+56$ days.     
   A variable point source was detected in all bands by the automated GROND pipeline 
   \citep{2008AIPC.1000..227Y}. The position of the transient 
   was calculated to be R.A. (J2000) = 04:00:42.62 and Dec (J2000) 
   = $-$55:57:20.0 compared to USNO-B reference field stars \citep{2003AJ....125..984M} 
   with an astrometric uncertainty of $0.\!\!^{\prime\prime}3$. 
   
   The optical and NIR image reduction and photometry were performed using standard
   IRAF tasks \citep{1993ASPC...52..173T} similar to the procedure described in detail
   in \citet{2008ApJ...685..376K}. A general model for the point-spread function 
   (PSF) of each image was constructed using bright field stars and fitted to the 
   afterglow. In addition, aperture photometry was carried out, and the results
   were consistent with the reported PSF photometry. All data were corrected
   for a Galactic foreground reddening of $E_{\mathrm{B-V}}=0.016$ mag in the direction 
   of the burst \citep{1998ApJ...500..525S}, corresponding to an extinction
   of $A_V=0.05$ using $R_V=3.1$.
   Optical photometric calibration was performed relative to the magnitudes of four secondary 
   standards in the GRB field, shown in Fig. \ref{chart091029} and Table \ref{standards091029}. During 
   photometric
   conditions, an SDSS field \citep{2002AJ....123.2121S} at R.A. (J2000) = 03:50:03.25, 
   Dec (J2000) = $-$00:00:37.9 was observed within 
   a few minutes of observations of the GRB field. The obtained zeropoints were corrected
   for atmospheric extinction and used to calibrate stars in the GRB 
   field. The apparent magnitudes of the afterglow were measured with respect to the secondary standards
   reported in Table \ref{standards091029}. The absolute calibration of the $JHK_s$ bands
   was obtained with respect to magnitudes of the Two Micron All Sky Survey
   (2MASS) stars within the GRB field obtained from the 2MASS catalog \citep{2006AJ....131.1163S} and converted to AB magnitudes. 
   All GROND data are listed in Tables \ref{091029griz} and \ref{091029JHK}.

 \subsection{Stardome and BOOTES-3/YA}
    The afterglow was observed with the Stardome 0.4~m telescope,
	located in Auckland (North Island, New Zealand), using
	a SBIG ST-L-6303E CCD. Images were obtained through a
	OG530 (Schott) filter \footnote{\url{http://www.optical-filters.co.uk/og530.html}} that transmits wavelengths above 5300 {\AA}ngstr\"om.
	The observations consisted of 300 s exposures that were combined
	in sets of 6 to improve the S/N ratio. 
	Further observations were obtained with the Yock-Allen (YA) robotic telescope at the
    BOOTES-3 astronomical station \citep{2011AcPol..51b..16C}, a fast-slewing 0.6 m Ritchey-Chr{\'e}tien telescope equipped with an
    iXon-889 EMCCD camera located in Blenheim (South Island, New Zealand). The
    observations consisted of a series of 60 s unfiltered exposures, which were combined
    in groups to improve the S/N ratio. Image reduction was done using standard techniques in IRAF. Aperture photometry was
    performed, using PHOT with apertures equivalent to the seeing. For Stardome images, a PSF-matching photometry was preferred. All data were then cross-calibrated using
    GROND photometry to obtain consistent results and are listed in Tables \ref{091029bootes} and \ref{091029stardome}.

  \begin{figure}[h]
    \resizebox{\hsize}{!}{\includegraphics{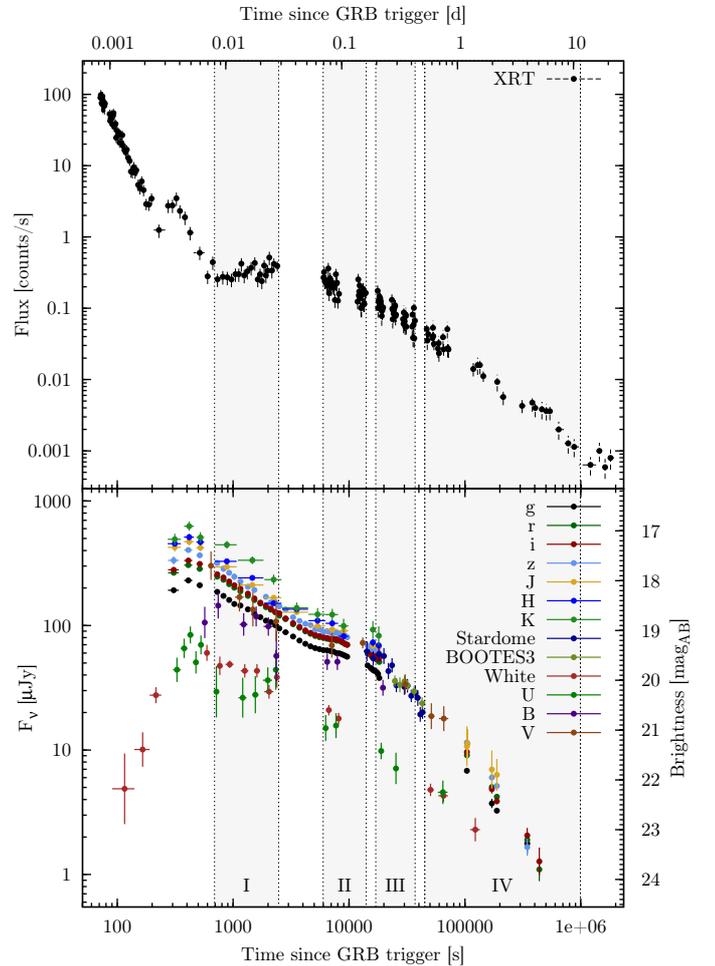}}
    \caption{Light curve of the X-ray (top panel) and ultraviolet, optical and near-infrared (bottom panel)
     afterglow of GRB 091029. Shown data are corrected for Galactic foreground extinction and
       are in AB magnitudes. Upper limits are not shown for better clarity.
       Gray regions show the time intervals where XRT data were obtained for the broad-band SEDs
      (Fig. \ref{091029BBplot}).}
    \label{091029lightcurve}
  \end{figure} 

\section{Results}
 \subsection{Afterglow light curve}
\begin{figure}[h]
    \resizebox{\hsize}{!}{\includegraphics[angle=270]{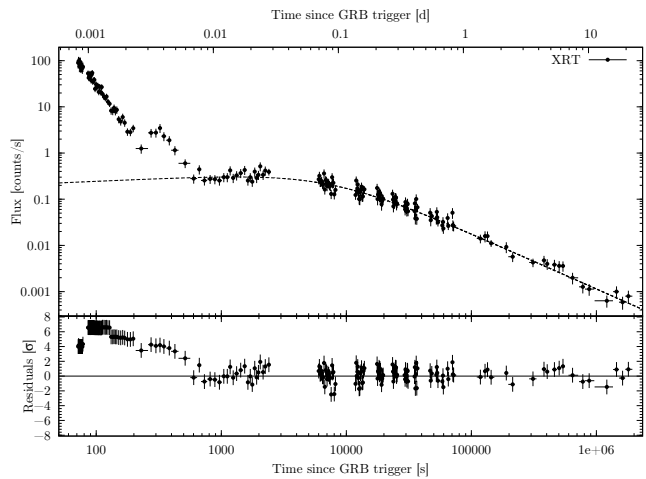}}
   \caption{The smoothly broken power-law fit to the X-ray light curve of the GRB 091029 afterglow. The fitting was applied to datapoints
             after 700~s in order to exclude the early steep transition phase and flaring.}
   \label{091029XRT}
 \end{figure}

   The X-ray light curve of the afterglow of GRB 091029 (Fig. \ref{091029lightcurve}) shows a very steep decay ($\alpha=3.69\pm0.10$) 
   until $\sim200$~s, consistent with being the tail of the GRB emission \citep{1996ApJ...473..998F}, 
   connecting the prompt phase of the GRB and its afterglow. 
   A steep X-ray flare follows after the decay  , 
   which declines rapidly with $\alpha=3.91\pm0.39$. 
   Given that the temporal decay indices before the flare and after its peak are consistent within $1\sigma$, 
   the possible scenario for this rapid rebrightening might involve a refreshed shock \citep{1998ApJ...496L...1R,1998ApJ...503..314P,2000ApJ...535L..33S,
   2000ApJ...532..286K,2002ApJ...566..712Z}, although the flare might be too rapid for this scenario \citep{2006ApJ...637..873H}.
   As we are predominantly interested in the afterglow phase of the GRB evolution, we exclude this flare from all our fits.
   The X-ray light curve after $\sim700$~s (Fig. \ref{091029XRT}) is best fitted with a broken power-law with a 
   smooth break \citep{1999A&A...352L..26B}. The best-fit (red. $\chi^2 = 0.89$) values of this model are $\alpha_{X1}=-0.12\pm0.10$, 
   $t_{\rm{break}}=7.4\pm1.8$~ks and $\alpha_{X2}=1.20\pm0.04$. The smoothness $s$ iterated to a value of 1, which was set as a lower bound in 
   the fit to better constrain the values of the temporal decay indices before and after the break.
    
  \begin{figure}[h]
    \resizebox{\hsize}{!}{\includegraphics[angle=270]{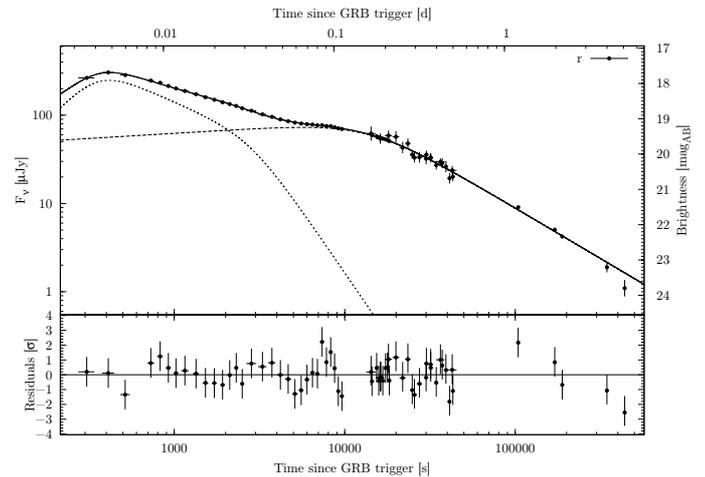}}
   \caption{The two-component fit to the $r'$-band data obtained by GROND, Stardome and BOOTES-3/YA. The parameters of the fit for both 
               components (dotted lines) are listed in Table \ref{091029rbandtable}. The solid line represents the superposition of the two
                components and the best fit to the data.}
   \label{091029rband}
 \end{figure}
   
   The optical/NIR light curve of the GRB 091029 afterglow shows a steep initial rise from the start of the observations 
   until the peak at around 400~s. The temporal slope of the rise, obtained from a fit of UVOT datapoints, 
   is $\alpha=-2.90 \pm 0.67$, consistent with the jet expanding in an ISM environment \citep{2008MNRAS.387..497P}. 
   The early peak in the 
   optical/NIR light curve is probably not the counterpart of the X-ray flare due to the time shift of both peaks. 
   The decay following the initial peak has a slope of $\alpha=0.58\pm 0.01$ until around 5~ks, when it starts to flatten. 
   This decay index is obtained from a simultaneous fit to the GROND datapoints between $0.6-5$~ks in all bands. 
      
   However, fitting this shallow decay phase in each optical/NIR band separately shows a steepening of the temporal index with 
   increasing wavelength of the GROND filters (see Table \ref{091029fittable}), suggesting that the afterglow gets bluer. 
   To fit the whole complex optical/NIR light curve from the beginning to the end of the observations, a two-component model is proposed.
   The first component, which dominates the observed optical light curve up until the bump at $T_0+5$~ks, is composed of three smoothly connected power-laws. The second component
   was needed to model the later hump and steep decay, and uses two smoothly connected power-laws. The obtained parameters of this fit (red. $\chi^2 = 0.92$)
   are listed in Table \ref{091029rbandtable} and are discussed later.
   
  \begin{table}[h]
   \caption{Light-curve fit parameters for the afterglow of GRB~091029 in the time interval of $0.6-5$~ks. 
   The fitting of the NIR bands is affected by the somewhat lower signal-to-noise ratio of the NIR data as compared 
   to the optical bands. The higher value of red. $\chi^2$ in the $H$ band is caused by larger residuals.}
   \label{091029fittable}      
   \centering                         
   \begin{tabular}{c c c}        
   \hline\hline    
   \\            
    Bands & $\alpha$ & $\chi^2$/d.o.f.\\    
    \hline 
    $g'r'i'z'JHK_s$ & $0.576 \pm 0.004$ & 68 / 71 \\
    $g'$ & $0.542 \pm 0.009$ & 2.9 / 15 \\
    $r'$ & $0.574 \pm 0.006$ & 5.1 / 15 \\
    $i'$ & $0.593 \pm 0.010$ & 3.2 / 15 \\
    $z'$ & $0.622 \pm 0.018$ & 6.1 / 14 \\
    $J$ & $0.601 \pm 0.028$ & 0.8 / 2 \\
    $H$ & $0.672 \pm 0.047$ & 9.6 / 2 \\
    $K_s$ & $0.815 \pm 0.075$ & 2.9 / 2 \\
    \hline
   \end{tabular}
   \end{table} 

  \begin{table*}[ht]
   \caption{Light-curve fit parameters for the whole set of $r'$-band data obtained by GROND, Stardome and BOOTES-3/YA.}
   \label{091029rbandtable}      
   \centering                        
   \begin{tabular}{c c c c c c c c}        
   \hline\hline    
   \\            
    $ F_\nu(t)$ & $\alpha_1$ & $t_1 [ks]$ & $s_1$ & $\alpha_2$ & $t_2 [ks]$ & $s_2$ & $\alpha_3$ \\      
    \hline 
    TPL$^{\mathrm{(a)}}$ & $-1.95$ (fixed) & $0.36\pm0.02$ & $2.0\pm0.1$ & $0.84\pm0.05$ & $2.80\pm0.15$ & $2.0\pm0.2$ & $2.83\pm0.20$ \\
    DPL$^{\mathrm{(b)}}$ & $-0.12\pm0.07$ & $13.90\pm0.80$ & $2.0\pm0.1$ & $1.14\pm0.02$ &  &  &  \\
    \hline
   \end{tabular}
   \begin{list}{}{}
    \item[$^{\mathrm{(a)}}$] Smoothly connected triple power-law, describing the narrow jet
    \item[$^{\mathrm{(b)}}$] Smoothly connected double power-law, describing the wider jet
    \end{list}
   \end{table*}  
   
 \subsection{Afterglow SEDs}
\begin{figure}[h]
   \resizebox{\hsize}{!}{\includegraphics{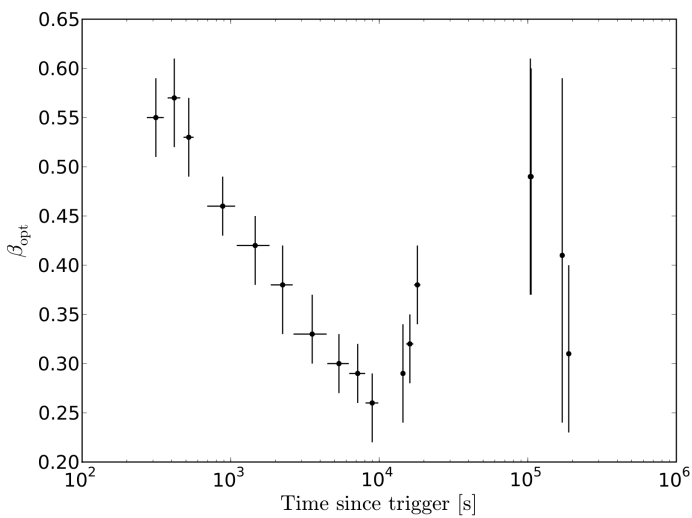}}
   \caption{The optical/NIR spectral slope as a function of time.}
   \label{091029sed}
 \end{figure}

   Given that the differences in decay slopes for each GROND filter point to a colour evolution, we need to study the SEDs of the afterglow. 
   Thanks to the simultaneous multi-band observing capabilities of GROND, it is possible to measure the spectral slope $\beta$ of the 
   optical/NIR data as a function of time. Fig. \ref{091029sed} shows that the optical/NIR spectral index decreases from 
   $0.57\pm0.04$ to $0.26\pm0.03$ between 0.4 and 9~ks, and then slowly increases again to a value of $0.49\pm0.12$ at around 100~ks.
   In addition, broad-band optical/NIR to X-ray SEDs were constructed at four different time intervals within this period, which are indicated in the 
   light curve (Fig. \ref{091029lightcurve}). Spectra were grouped using the grppha task and fitted with the GROND data in XSPEC v12 using 
   $\chi^2$ statistics. The combined optical/X-ray SEDs were fitted with power-law 
   and broken power-law models and two absorbing columns: one Galactic foreground with a hydrogen 
   column of $N_H = 1.14\times10^{20}$~cm$^{-2}$ \citep{2005A&A...440..775K} and another one that is 
   local to the GRB host galaxy at $z=2.75$. 
   Only the latter was allowed to vary in the fits. To investigate the dust 
   reddening in the GRB environment, the zdust model was used, which contains Large and Small Magellanic Clouds 
   (LMC, SMC) and Milky Way (MW) extinction laws from \citet{1992ApJ...395..130P}.
   Fits of optical/NIR data alone as well as the broad-band fits 
   resulted in a host dust extinction that was consistent with zero, therefore in all the models we 
   assumed no host dust extinction for simplicity. 
   With photometric data alone it was not possible to constrain the presence of Lyman-alpha absorption \citep{2000ApJ...536....1L} in the $g'$ band from 
   neutral hydrogen within the host galaxy. The $g'$-band data were therefore removed from the SED fits. 
   
  \begin{figure}[h]
    \resizebox{\hsize}{!}{\includegraphics[angle=270]{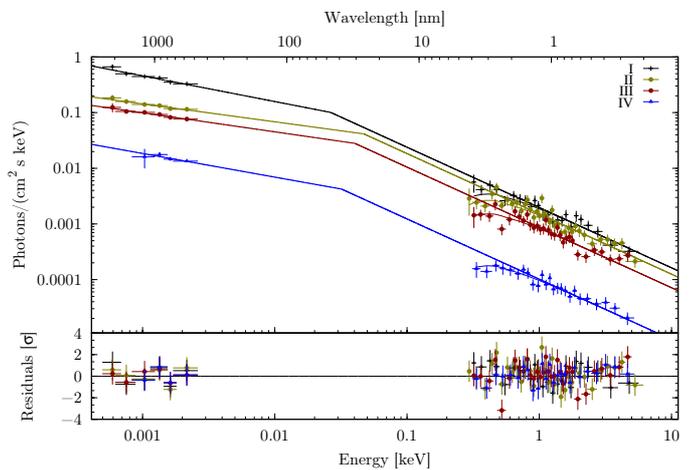}}
   \caption{Broad-band optical/NIR to X-ray SEDs fitted with a broken power-law. Mid-times of each SED are listed in Table \ref{091029BBtable} and 
             regions where the X-ray data were taken for each SED are shown in Fig. \ref{091029lightcurve}.}
   \label{091029BBplot}
 \end{figure}
   
   Given that the broad-band SEDs proved to be inconsistent with a simple power-law model (red. $\chi^2 = 16.5$), we used models
   that include a break between the X-ray and optical/NIR data. 
   We fitted all four epochs of broad-band SEDs simultaneously with a sharp broken power-law model,
   where the host-intrinsic absorbing column density and the X-ray spectral index are tied between each SED but left
   free to vary (Fig. \ref{091029BBplot}), due to the fact that the X-ray hardness ratio does not change during the afterglow. 
   The low-energy spectral indices and energy of the break were left untied between SEDs and free to vary.
   The best fit (red. $\chi^2 = 0.94$) gives values of the host-equivalent neutral hydrogen 
   density $N_{H,X} = (3.0\pm1.3)\times10^{21}$~cm$^{-2}$
   and a high-energy spectral index $\beta_X=1.08^{+0.06}_{-0.05}$. The best-fit values of the low-energy spectral indices and break energies are listed 
   in Table \ref{091029BBtable}. This fit shows that the break evolves in time to shorter wavelengths between SEDs I and II, and then it evolves
   the opposite way towards longer wavelengths between SEDs II and IV.  It also shows that below and above the cooling break $\Delta\beta \neq 0.5$ 
   (with quite high and variable significance), which is a value demanded by theory \citep{1998ApJ...497L..17S}. This is
   another indication that we are not seeing a simple single radiating electron population.
    
 \begin{table}[ht]
   \caption{Best-fit parameters resulting from the broken power-law fit to the broad-band SEDs. 
            The host-equivalent neutral hydrogen density $N_H = (3.0\pm1.3)\times10^{21}$~cm$^{-2}$.}
   \label{091029BBtable}      
   \centering                           
   \begin{tabular}{c c c c c}        
   \hline\hline    
   \\            
    SED & Midtime [s] & Low energy & Cooling & High energy  \\  
    number & of SED & spectral index & break [eV] & spectral index \\
    \hline \\
    I & 883 & $0.46^{+0.06}_{-0.06}$ & $26.4^{+15.2}_{-9.8}$ & \multirow{4}{*}{$1.08^{+0.06}_{-0.05}$} \\ [6pt]
    II & 7161 & $0.32^{+0.05}_{-0.06}$ & $47.2^{+20.8}_{-14.5}$ & \\ [6pt]
    III & 18056 & $0.34^{+0.06}_{-0.06}$ & $40.4^{+19.9}_{-13.1}$ &  \\ [6pt]
    IV & 104026 & $0.42^{+0.23}_{-0.21}$ & $31.6^{+115.5}_{-17.0}$ & \\ 
    \hline
   \end{tabular}
  \end{table}
 
\subsection{Closure relations} 
   Using values obtained from the different fits we can test the closure relations
   \citep{2002ApJ...568..820G,2001ApJ...558L.109D,2004IJMPA..19.2385Z,2009ApJ...698...43R,2000ApJ...543...66P}
   between temporal and spectral indices. Assuming the break in the broad-band SEDs to be the cooling break $\nu_c$, we see that the X-ray data
   are above this frequency and the optical/NIR data below it during the afterglow observations. 
   The fit-derived X-ray spectral index $\beta_X=1.08^{+0.06}_{-0.05}$ results in a 
   power-law index of the electron energy distribution $p = 2.17 \pm 0.11$. This spectral index and the late X-ray temporal slope 
   of $\alpha_{X2}=1.20\pm0.04$ are within $1\sigma$ consistent with the equation \citep{2009ApJ...698...43R} for $p>2$ and a constant decay in the 
   $\nu > \nu_c$ regime, where the jet is interacting with either a homogeneous interstellar medium (ISM) or a wind and is in the slow or fast 
   cooling phase. 

   The late ($t>20$~ks) optical/NIR single-component model decay index of $\alpha=1.14\pm0.02$ and the late spectral index of $\beta=0.49\pm0.12$ are 
   within $1\sigma$ consistent with the equation for a normal decay in the $\nu_m < \nu < \nu_c$ regime, where the jet is interacting
   with a wind medium and is in the slow cooling phase. However, the steep initial onset of the optical/NIR afterglow seems to exclude this scenario, as it is well 
   consistent with the expected temporal index of $\alpha \sim -3$ for the jet interacting with the ISM and $\nu_c > \nu_{opt}$,
   whereas the fastest possible rise for a wind medium is $\alpha \sim -0.5$ \citep{2008MNRAS.387..497P}. 
   Moreover, during the early shallow decay in the optical/NIR bands, the spectral index
   is evolving with time and thus cannot be tested with the simple closure relations. 
   In the case of the two-component scenario (Fig. \ref{091029rband}), the resulting spectral index $\beta$ is the superposition of the 
   spectral indices of the narrow and the wide jet and evolves with time as the ratio between these two jets changes. Without knowing the
   spectral indices of each component, the closure relations for such a scenario cannot be tested. 
   
\section{Discussion}
   The almost total decoupling of the optical/NIR and the X-ray light curves of the afterglow of GRB~091029 suggests a double outflow origin
   \citep{2004ApJ...605..300H,2005ApJ...626..966P,2003ApJ...595L..33S,2007ApJ...656L..57J,2005MNRAS.357.1197W,2005A&A...440..477R,2008Natur.455..183R,2011A&A...526A.113F}. 
   This is supported by our finding that the X-ray spectral hardness does not evolve synchronously with the optical spectral
   hardening at $0.3-10$~ks. We discuss three possible scenarios, all based on a two-component jet, to explain the peculiar behavior of the light
   curve and the spectrum of this afterglow. 
   
 \subsection{Continuous energy injection}
   In this scenario, the X-ray light curve after $700$~s is produced by the wider, X-ray-dominating outflow, which has a deceleration 
   time of a few ks. The pre-deceleration phase of the wide jet would cause the early shallow rise of the X-ray afterglow with the peak around $7.4$~ks, after 
   which the wide jet would turn into normal deceleration producing the $\alpha_{X2} \sim 1.2$ decay. The same principle would apply
   to the early optical light curve, where the early rise and peak at $\sim 400$~s would be a result of a pre-deceleration phase of the narrower, 
   optically dominating outflow. From the peak time of this narrower outflow, we can estimate the initial Lorentz factor in case of the ISM to be $\Gamma_n \sim 300$ using 
   \citet{2007A&A...469L..13M} or $\Gamma_n \sim 120$ using \citet{2012MNRAS.420..483G}, in both cases
   substituting the measured quantities and normalizing to the typical values $n = 1$~cm$^{-3}$ and $\eta = 0.2$ \citep{2003ApJ...594..674B}.
   The shallow decay of the optical/NIR light curve until 10~ks is then a result of some form of a continuous energy 
   injection by the central engine \citep{1998ApJ...496L...1R,1998PhRvL..81.4301D,1998ApJ...503..314P,2000ApJ...535L..33S,2001ApJ...552L..35Z,2006ApJ...642..354Z,2011A&A...529A.142R}.
   When this energy injection in the narrow jet ends at $\sim20$~ks, the temporal slope steepens to $\alpha \sim 1.1$, characteristic for a normal 
   decay. However, this scenario cannot explain the time evolution of the optical/NIR spectral index during the injection time interval. 
   The hardening of the optical spectrum would require that the electron index $p$ changes in the outflow with time and after the injection 
   ends, it changes back close to its original value (Fig. \ref{091029sed}), making this scenario somewhat contrived.
   
   Another issue with this scenario is the fact that in the standard interpretations, the narrow jet dominates the X-ray afterglow as it is more
   energetic, and the wider jet dominates the optical domain, especially at later times. The X-ray light curve is therefore expected to peak before the optical one. 
   In such a standard interpretation, the reverse order of peaks 
   in the light curves of the afterglow of GRB~091029 could be the result of an off-axis geometry of the jet as seen by the observer. If the observer
   was inside the cone of the wide, optically dominating jet, but outside of the beaming cone of the narrow, X-ray dominating jet, the optical 
   light curve could start with a steep rise and early peak due to the pre-deceleration phase of the wide jet, and the X-ray light curve could follow 
   with a later peak as the relativistically beamed emission cone of the narrow jet widens and gradually enters the sightline of the observer.
   However, given that neither the early shallow decay of the optical/NIR light curve nor the colour evolution in the same domain are correlated with the temporal 
   evolution of the narrow jet, this off-axis scenario has the same problems with the unfeasible temporal evolution of the electron energy distribution
   as the previous one.
   
 \subsection{Two outflows with different $p$ parameters}
   The second scenario uses a different two-component model, without energy injection, where the optical/NIR light curve is a superposition of two components
   as described in section 3.1 and shown in Fig. \ref{091029rband}. The first component would represent a narrow, 
   ultra-relativistic jet, with deceleration time of $\sim 400$~s, a normal decay phase afterwards, and a jet break at $\sim 2.8$~ks, followed by
   a steep post-jet-break decay. The second component represents a wider and mildly relativistic jet with the deceleration time of $\sim 14$~ks and
   a normal decay. 
   While the residuals in Fig. \ref{091029rband} might hint at a possible jet break of the wider component around 100~ks, the data are
   within $3\sigma$ of the straight power-law and there is no visible break in the X-ray data at that time. Therefore we cannot make a reliable statement
   about the presence of such a jet break. 
   The narrow jet would be dominant in the early part of the light curve and the 
   wide one would be responsible for the late hump and dominate the optical/NIR light curve afterwards. The shallow decay between $0.4-5$~ks
   would be a result of the superposition of fluxes from both outflows. Since the emission from the narrow component in the normal decay phase is decreasing
   and the the emission from the wide component in the pre-deceleration phase is increasing with time, the ratio of the fluxes of the two outflows in this time interval would vary.
   This model can explain the flattening of the spectral index during this
   period, assuming that each outflow with different Lorentz factors has a different electron energy distribution index $p$. As the ratio between 
   the narrow jet with a higher $p$ value ($p\sim2.1$, estimated from the highest value of the optical/NIR spectral slope) 
   and the wide jet with a lower $p$ value ($p\sim1.5$, estimated from the lowest value of the optical/NIR spectral slope) changes, the spectral index of the
   co-added flux is gradually evolving as well. The spectral index from the start of the GROND data is governed by the narrow outflow with the soft spectrum but gradually 
   decreases as the flux from the wide component with the harder spectrum gets dominant (Fig. \ref{091029rband}).
   However, this model does not explain the turnover of the spectral evolution at $\sim10$~ks, where
   only the wide component with a constant $p$ is dominant. 
   The model also does not fit the shape of the X-ray light curve, which should be dominated the whole time by the narrow jet due to the X-ray
   spectral index $\beta_X$ being constant and consistent with $p\sim2.1$ of the narrow jet.  
   
 \subsection{Two outflows in different stages of the spectral evolution}
   The third scenario uses a slightly modified two-component jet setup described in section 4.2 (Fig. \ref{091029rband}), in which now both outflows have the same value 
   of $p\sim2.1$ and are at different stages of the synchrotron spectral evolution \citep{1998ApJ...497L..17S}. Both the X-ray and optical/NIR light curves
   can be modeled as superpositions of the two components. The flattening of the SEDs II and III in 
   the optical/NIR region (Fig. \ref{091029BBplot}) would then be a result of the wide jet having both the cooling break $\nu_c$ and the injection 
   frequency $\nu_m$ between the X-ray and optical/NIR wavelengths \citep{2012A&A...538L...7N}, while the narrow jet has only the cooling frequency $\nu_c$ between X-ray and optical/NIR bands.
   In that case, the spectral slope in the optical/NIR bands of the narrow
   jet would be $\beta_X-0.5=0.58\pm0.06$ and that of the wide jet would be $-1/3$ \citep{1998ApJ...497L..17S}. As the ratio between these two outflows  
   changes, it would explain the spectral hardening in the optical/NIR bands, while being consistent with the X-ray spectral slope staying 
   constant thanks to equal $p$ values in both outflows. The turnover in the spectral evolution at $\sim10$~ks can be explained by the passage
   of the frequency $\nu_m$ through the GROND filters, after which the optical/NIR spectral index would be consistent with the spectral phase of the narrow jet. 
   Given that the softening of the optical/NIR spectrum after $\sim10$~ks is slow and gradual, the spectral break at the frequency $\nu_m$ must
   be very smooth \citep{2002ApJ...568..820G,2011A&A...535A..57F}. 
   
   This model can also explain different break times in the X-ray and optical/NIR
   light curves, assuming that the deceleration time of the wide jet is defined by the X-ray light-curve peak but the movement of the break $\nu_m$ is 
   counter-balancing the expected decay steepening in the optical/NIR bands until it passes through them and causes the late break in the light curve at $T_0 = 10$~ks.
   Of course, it is questionable whether the movement of the break $\nu_m$ could counter-balance the turnover visible in the X-ray light curve so perfectly that it would
   in fact completely negate it in the optical/NIR domain and produce such a straight power-law as we see in the optical/NIR light curve between $0.4-5$~ks.  
   This scenario is difficult to confirm or disprove, though, by fitting the light curve and SEDs alone because this model has a large number of
   free parameters. Therefore it is very difficult to fit the broad-band SEDs with a model consisting of a double power-law and a triple power-law 
   component and constrain all the spectral slopes and synchrotron break frequencies reliably. The fitting of light curves is confronted with similar
   difficulties. The optical/NIR light curve might need an even more complex model than the one presented in Fig. \ref{091029rband} in order to describe the effects
   of the moving frequency $\nu_m$. However, as our data are the result of the superposition of two components, fitting of the data does not provide us with reliable constraints on 
   the parameters of each component.
   
\section{Conclusions}
   The growing number of well-sampled data sets from the latest generation of instruments like the \emph{Swift} satellite and the GROND imager 
   show that the radiative mechanism responsible for the optical to X-ray GRB afterglow emission is not as simple and well  
   understood as previously believed. The simplest fireball model has an increasingly difficult time to explain the complex light curves
   of some GRB afterglows. In many cases, the optical and X-ray emission are seemingly decoupled, thus providing an indication that they 
   are produced by different mechanisms. The afterglow of GRB 091029 is an extreme case, where the optical/NIR and the X-ray light curves are 
   almost totally decoupled, as if they belonged to two different GRBs. Moreover, the GROND SEDs show a strong colour evolution with the optical/NIR 
   spectral index decreasing from $0.57$ to $0.26$ between 0.4 and 9~ks, and then increasing again to a value of $\sim0.49$ at around 100~ks, 
   while the X-ray spectral index remains constant throughout the observations. This observational evidence leads us to the conclusion that the emission 
   in both energy bands needs to be produced by two different outflows.
   
   We discuss several possible scenarios to explain this peculiar afterglow. The first one includes a continuous energy injection to explain
   the shallow initial decay of the optical/NIR light curve. However, this model is not able to explain the spectral evolution during the injection
   period, given that the theory assumes the electron energy distribution index $p$ of the outflow is constant. To solve this, the second scenario
   uses two components with different $p$ values. As the ratio between these two outflows changes, the resulting spectral index changes as well. 
   This model is, however, not able to explain either the turnover in the optical/NIR spectral-slope evolution, nor the different times of the breaks in 
   the X-ray and optical/NIR light curves. The third scenario offers a solution by putting the two outflows with similar $p$ values into two different 
   stages of the spectral evolution. The narrow jet, dominating the optical/NIR wavelengths before the hump, has a cooling break between the optical
   and the X-ray bands, while the wide jet, responsible for the late optical/NIR light curve, has both $\nu_c$ and $\nu_m$ frequencies
   between the optical and the X-ray bands. During the hump, the injection frequency $\nu_m$ passes through the GROND filters and the light curve becomes
   similar to the X-ray one. 
   
   Even though the last model can in principle explain the irregularities in the afterglow of GRB~091029, its complexity does not allow us to test
   it reliably, despite the large high-quality dataset presented in this work. 
   Ironically, the data quality of the presented afterglow light curve is so good it rules out any simple model for the temporal-spectral evolution,
   but is not good enough to really constrain the more complex, constructed models.
   Rather than using the forward-shock scenario,  alternative models might be needed
   to explain the multi-wavelength data of the afterglow of GRB 091029. For example a reverse-forward shock emission, where the optical afterglow is dominated by a long-lived reverse
   shock \citep{2007ApJ...665L..93U,2007MNRAS.381..732G} and the X-ray afterglow is from the forward shock. This model can decouple the two light curves almost completely, requires mass injection into the blast-wave
   and has many free parameters. Another alternative can be the cannonball model \citep{2010ApJ...712.1172D}, where the observed afterglow emission is
   described as the sum of thermal bremsstrahlung and synchrotron emission produced by one or more cannonballs decelerating in the circumburst medium. The last we mention is the late prompt
   model \citep{2007ApJ...658L..75G,2008MNRAS.388.1729K,2008Sci...321..376K,2009MNRAS.393..253G,2010MNRAS.403.1131N}, which was proposed to explain the 
   different temporal and spectral evolution in the optical and X-ray bands of certain long
   GRBs which show a late-time chromatic flattening. The authors interpret the complex broadband evolution as due to the sum of two separate processes: the standard
   forward shock and the emission produced by a late-time activity of the central engine (i.e., the so called late prompt emission).
   
\begin{acknowledgements}
We thank the anonymous referee for constructive comments
that helped to improve the paper.\\
TK and DAK acknowledge support by the DFG cluster of excellence Origin and 
Structure of the Universe.\\
TK acknowledges support by the European Commission under the Marie Curie
Intra-European Fellowship Programme.\\
The Dark Cosmology Centre is funded by the Danish National Research Foundation.\\
FOE acknowledges funding of his Ph.D. through the \emph{Deutscher Akademischer Austausch-Dienst} (DAAD).\\
SK, AR, DAK and ANG acknowledge support by DFG grant Kl 766/16-1.\\
AR acknowledges support from the {\it Jenaer Graduierten\-akademie}.\\
MN and PS acknowledge support by DFG grant SA 2001/2-1.\\
ANG, DAK and AR are grateful for travel funding support through MPE.\\
SRO acknowledges support from the UK Space Agency.\\
Work by SD was performed under contract with the California Institute
of Technology (Caltech) funded by NASA through the Sagan Fellowship
Program.\\
We acknowledge support from
the Spanish Ministry through project AYA 2009-14000-C03-01 (including FEDER
Funds).\\
GROND: Part of the funding for GROND (both hardware as well as 
personnel) was generously granted from the Leibniz-Prize to Prof. G. 
Hasinger (DFG grant HA 1850/28-1).\\
\emph{Swift}: This work made use of data supplied by the UK \emph{Swift} Science 
Data Centre at the University of Leicester.
\end{acknowledgements} 

\bibliographystyle{aa}
\bibliography{biblio_thesis}

\newpage
\begin{longtable}{r c c c}
\caption{\label{091029uvot} UVOT photometric data}   \\          
\hline\hline
$T_\mathrm{mid} - T_0$ [ks] & Exposure [s] & Brightness$^{\mathrm{(a)}}$ mag$_\mathrm{AB}$ & Filter\\
\hline
\endfirsthead
\caption{continued.}\\
\hline\hline
$T_\mathrm{mid} - T_0$ [ks] & Exposure [s] & Brightness$^{\mathrm{(a)}}$ mag$_\mathrm{AB}$ & Filter\\
\hline
\endhead
\hline  
0.1156	&	50	&$	22.18	\pm	0.71	$&$	White	$\\
0.1656	&	50	&$	21.39	\pm	0.35	$&$	White	$\\
0.2155	&	50	&$	20.30	\pm	0.16	$&$	White	$\\
0.5932	&	20	&$	19.45	\pm	0.16	$&$	White	$\\
0.7674	&	20	&$	19.71	\pm	0.19	$&$	White	$\\
0.9332	&	150	&$	19.68	\pm	0.08	$&$	White	$\\
1.2584	&	193	&$	19.81	\pm	0.14	$&$	White	$\\
1.6049	&	193	&$	19.81	\pm	0.14	$&$	White	$\\
2.0387	&	366	&$	20.23	\pm	0.14	$&$	White	$\\
2.3850	&	20	&$	19.94	\pm	0.21	$&$	White	$\\
6.6930	&	200	&$	20.60	\pm	0.11	$&$	White	$\\
8.1318	&	200	&$	20.77	\pm	0.12	$&$	White	$\\
50.5180	&	6487	&$	22.20	\pm	0.13	$&$	White	$\\
65.1696	&	12382	&$	22.32	\pm	0.13	$&$	White	$\\
122.7112	&	23679	&$	23.00	\pm	0.24	$&$	White	$\\
146.0998	&	12138	&$	>22.94			$&$	White	$\\
203.9169	&	35415	&$	>23.06			$&$	White	$\\
311.1750	&	40962	&$	>23.11			$&$	White	$\\
404.1329	&	75705	&$	>23.4			$&$	White	$\\
496.6075	&	98012	&$	>23.91			$&$	White	$\\
554.1967	&	6354	&$	>22.91			$&$	White	$\\
0.3277	&	50	&$	19.79	\pm	0.25	$&$	U	$\\
0.3777	&	50	&$	19.36	\pm	0.20	$&$	U	$\\
0.4277	&	50	&$	19.09	\pm	0.17	$&$	U	$\\
0.4777	&	50	&$	19.64	\pm	0.23	$&$	U	$\\
0.5276	&	50	&$	19.29	\pm	0.19	$&$	U	$\\
0.7181	&	20	&$	20.23	\pm	0.52	$&$	U	$\\
1.2085	&	193	&$	20.35	\pm	0.40	$&$	U	$\\
1.5553	&	193	&$	20.29	\pm	0.38	$&$	U	$\\
1.9896	&	366	&$	20.00	\pm	0.26	$&$	U	$\\
2.3351	&	20	&$	19.79	\pm	0.40	$&$	U	$\\
6.2829	&	200	&$	20.96	\pm	0.26	$&$	U	$\\
7.7213	&	200	&$	20.91	\pm	0.25	$&$	U	$\\
18.9286	&	908	&$	21.42	\pm	0.17	$&$	U	$\\
25.3918	&	444	&$	21.77	\pm	0.32	$&$	U	$\\
36.9765	&	414	&$	>21.93			$&$	U	$\\
49.8063	&	6486	&$	>22.07			$&$	U	$\\
64.5991	&	12538	&$	22.25	\pm	0.24	$&$	U	$\\
75.6289	&	337	&$	>21.39			$&$	U	$\\
122.3652	&	23602	&$	>22.96			$&$	U	$\\
145.6560	&	12214	&$	>22.10			$&$	U	$\\
203.5215	&	35436	&$	>22.73			$&$	U	$\\
310.5329	&	40872	&$	>23.20			$&$	U	$\\
403.6852	&	75885	&$	>22.98			$&$	U	$\\
473.2423	&	52055	&$	>22.84			$&$	U	$\\
635.9086	&	75370	&$	>23.60			$&$	U	$\\
985.0417	&	47426	&$	>23.42			$&$	U	$\\
1340.8648	&	52778	&$	>23.74			$&$	U	$\\
1669.9707	&	88045	&$	>23.81			$&$	U	$\\
2034.5151	&	41876	&$	>23.66			$&$	U	$\\
0.5689	&	20	&$	18.84	\pm	0.32	$&$	B	$\\
0.7431	&	20	&$	18.50	\pm	0.26	$&$	B	$\\
1.2336	&	194	&$	18.88	\pm	0.23	$&$	B	$\\
1.5803	&	193	&$	18.72	\pm	0.20	$&$	B	$\\
2.0143	&	366	&$	18.92	\pm	0.19	$&$	B	$\\
2.3598	&	20	&$	19.51	\pm	0.51	$&$	B	$\\
6.4882	&	200	&$	19.63	\pm	0.17	$&$	B	$\\
7.9271	&	200	&$	19.63	\pm	0.17	$&$	B	$\\
19.6085	&	441	&$	20.15	\pm	0.16	$&$	B	$\\
310.8542	&	40917	&$	>21.78			$&$	B	$\\
403.9090	&	75794	&$	>22.88			$&$	B	$\\
473.4810	&	52145	&$	>23.00			$&$	B	$\\
0.0745	&	10	&$	>18.19			$&$	V	$\\
0.6446	&	20	&$	17.70	\pm	0.29	$&$	V	$\\
0.8170	&	20	&$	>17.85			$&$	V	$\\
1.1348	&	193	&$	18.33	\pm	0.29	$&$	V	$\\
1.4813	&	192	&$	18.58	\pm	0.34	$&$	V	$\\
1.9159	&	366	&$	>18.59			$&$	V	$\\
2.3483	&	193	&$	18.82	\pm	0.40	$&$	V	$\\
7.1045	&	200	&$	19.30	\pm	0.25	$&$	V	$\\
13.1297	&	907	&$	19.25	\pm	0.11	$&$	V	$\\
30.4949	&	907	&$	20.08	\pm	0.20	$&$	V	$\\
51.2643	&	6556	&$	20.72	\pm	0.26	$&$	V	$\\
65.7636	&	12270	&$	20.77	\pm	0.25	$&$	V	$\\
123.0766	&	23794	&$	>21.18			$&$	V	$\\
146.5634	&	12101	&$	>21.11			$&$	V	$\\
204.3240	&	35420	&$	>21.77			$&$	V	$\\
311.4738	&	40960	&$	>21.01			$&$	V	$\\
407.2422	&	81383	&$	>20.92			$&$	V	$\\
476.8197	&	46508	&$	>20.98			$&$	V	$\\
\hline
\end{longtable}
\begin{list}{}{}
\item[$^{\mathrm{(a)}}$] Corrected for Galactic foreground reddening.
\end{list}

\begin{table*}
\caption{GROND $g'r'i'z'$ photometric data}             
\label{091029griz}      
\centering                         
\begin{tabular}{r c c c c c}        
\hline\hline    
\\            
$T_\mathrm{mid} - T_0$ [ks] & Exposure [s] & \multicolumn{4}{c}{
Brightness$^{\mathrm{(a)}}$ mag$_\mathrm{AB}$} \\
 & & $g'$ & $r'$ & $i'$ & $z'$ \\
\hline   
0.3067	&	66	&$	18.19	\pm	0.04	$&$	17.84	\pm	0.02	$&$	17.78	\pm	0.03	$&$	17.59	\pm	0.08	$\\
0.4103	&	66	&$	18.00	\pm	0.04	$&$	17.69	\pm	0.02	$&$	17.60	\pm	0.02	$&$	17.38	\pm	0.04	$\\
0.5154	&	66	&$	18.09	\pm	0.03	$&$	17.76	\pm	0.02	$&$	17.66	\pm	0.03	$&$	17.49	\pm	0.05	$\\
0.7290	&	66	&$	18.22	\pm	0.03	$&$	17.92	\pm	0.02	$&$	17.87	\pm	0.04	$&$	17.64	\pm	0.05	$\\
0.8257	&	66	&$	18.31	\pm	0.03	$&$	17.98	\pm	0.02	$&$	17.94	\pm	0.03	$&$	17.76	\pm	0.06	$\\
0.9237	&	66	&$	18.39	\pm	0.03	$&$	18.07	\pm	0.02	$&$	18.02	\pm	0.03	$&$	17.84	\pm	0.05	$\\
1.0238	&	66	&$	18.47	\pm	0.04	$&$	18.14	\pm	0.02	$&$	18.09	\pm	0.03	$&$	17.92	\pm	0.07	$\\
1.1549	&	115	&$	18.50	\pm	0.03	$&$	18.21	\pm	0.02	$&$	18.16	\pm	0.04	$&$	18.02	\pm	0.05	$\\
1.3407	&	115	&$	18.58	\pm	0.04	$&$	18.31	\pm	0.01	$&$	18.26	\pm	0.03	$&$	18.12	\pm	0.05	$\\
1.5249	&	115	&$	18.67	\pm	0.03	$&$	18.39	\pm	0.02	$&$	18.36	\pm	0.02	$&$	18.19	\pm	0.05	$\\
1.7169	&	115	&$	18.73	\pm	0.02	$&$	18.46	\pm	0.01	$&$	18.44	\pm	0.03	$&$	-			$\\
1.9195	&	115	&$	18.80	\pm	0.03	$&$	18.53	\pm	0.02	$&$	18.50	\pm	0.03	$&$	18.32	\pm	0.05	$\\
2.1132	&	115	&$	18.83	\pm	0.03	$&$	18.58	\pm	0.02	$&$	18.56	\pm	0.02	$&$	18.44	\pm	0.04	$\\
2.3095	&	115	&$	18.90	\pm	0.03	$&$	18.64	\pm	0.01	$&$	18.63	\pm	0.03	$&$	18.43	\pm	0.05	$\\
2.5040	&	115	&$	18.96	\pm	0.03	$&$	18.70	\pm	0.01	$&$	18.66	\pm	0.02	$&$	18.51	\pm	0.05	$\\
2.8368	&	375	&$	19.04	\pm	0.02	$&$	18.77	\pm	0.01	$&$	18.76	\pm	0.02	$&$	18.58	\pm	0.03	$\\
3.2896	&	375	&$	19.12	\pm	0.02	$&$	18.88	\pm	0.01	$&$	18.85	\pm	0.01	$&$	18.73	\pm	0.03	$\\
3.7418	&	375	&$	19.20	\pm	0.02	$&$	18.95	\pm	0.02	$&$	18.93	\pm	0.02	$&$	18.81	\pm	0.03	$\\
4.1939	&	375	&$	19.27	\pm	0.02	$&$	19.02	\pm	0.01	$&$	19.00	\pm	0.02	$&$	18.86	\pm	0.04	$\\
4.6568	&	375	&$	19.32	\pm	0.02	$&$	19.07	\pm	0.01	$&$	19.05	\pm	0.02	$&$	18.89	\pm	0.03	$\\
5.1021	&	375	&$	19.35	\pm	0.02	$&$	19.11	\pm	0.01	$&$	19.09	\pm	0.02	$&$	18.98	\pm	0.04	$\\
5.5542	&	375	&$	19.38	\pm	0.01	$&$	19.14	\pm	0.01	$&$	19.11	\pm	0.02	$&$	18.98	\pm	0.03	$\\
6.0063	&	375	&$	19.39	\pm	0.02	$&$	19.15	\pm	0.01	$&$	19.14	\pm	0.02	$&$	19.02	\pm	0.04	$\\
6.4708	&	375	&$	19.40	\pm	0.02	$&$	19.17	\pm	0.01	$&$	19.15	\pm	0.02	$&$	19.01	\pm	0.03	$\\
6.9143	&	375	&$	19.41	\pm	0.02	$&$	19.18	\pm	0.01	$&$	19.17	\pm	0.02	$&$	19.04	\pm	0.03	$\\
7.3624	&	375	&$	19.42	\pm	0.01	$&$	19.18	\pm	0.01	$&$	19.17	\pm	0.02	$&$	19.06	\pm	0.03	$\\
7.8060	&	375	&$	19.45	\pm	0.01	$&$	19.21	\pm	0.01	$&$	19.18	\pm	0.02	$&$	19.05	\pm	0.03	$\\
8.2685	&	375	&$	19.46	\pm	0.02	$&$	19.21	\pm	0.01	$&$	19.20	\pm	0.02	$&$	19.07	\pm	0.04	$\\
8.7202	&	375	&$	19.48	\pm	0.02	$&$	19.24	\pm	0.01	$&$	19.23	\pm	0.02	$&$	19.09	\pm	0.03	$\\
9.1693	&	375	&$	19.50	\pm	0.02	$&$	19.27	\pm	0.01	$&$	19.26	\pm	0.02	$&$	19.10	\pm	0.04	$\\
9.6214	&	375	&$	19.53	\pm	0.01	$&$	19.29	\pm	0.01	$&$	19.29	\pm	0.02	$&$	19.14	\pm	0.03	$\\
14.4446	&	375	&$	19.70	\pm	0.01	$&$	19.47	\pm	0.01	$&$	19.43	\pm	0.02	$&$	19.30	\pm	0.04	$\\
15.3777	&	375	&$	19.75	\pm	0.01	$&$	19.50	\pm	0.01	$&$	19.48	\pm	0.02	$&$	19.32	\pm	0.05	$\\
15.8274	&	375	&$	19.78	\pm	0.01	$&$	19.53	\pm	0.01	$&$	19.51	\pm	0.02	$&$	19.38	\pm	0.03	$\\
16.2785	&	375	&$	19.80	\pm	0.01	$&$	19.54	\pm	0.01	$&$	19.52	\pm	0.02	$&$	19.39	\pm	0.04	$\\
16.7300	&	375	&$	19.81	\pm	0.01	$&$	19.57	\pm	0.01	$&$	19.50	\pm	0.02	$&$	19.36	\pm	0.04	$\\
17.3570	&	375	&$	19.84	\pm	0.02	$&$	19.58	\pm	0.01	$&$	19.55	\pm	0.02	$&$	19.41	\pm	0.04	$\\
17.8057	&	375	&$	19.87	\pm	0.02	$&$	19.60	\pm	0.01	$&$	19.58	\pm	0.02	$&$	19.42	\pm	0.04	$\\
18.2565	&	375	&$	19.95	\pm	0.03	$&$	19.63	\pm	0.01	$&$	19.55	\pm	0.04	$&$	19.46	\pm	0.05	$\\
104.0014	&	686	&$	21.82	\pm	0.05	$&$	21.51	\pm	0.03	$&$	21.44	\pm	0.06	$&$	21.25	\pm	0.09	$\\
170.5107	&	1714	&$	22.47	\pm	0.09	$&$	22.14	\pm	0.06	$&$	22.19	\pm	0.09	$&$	21.95	\pm	0.11	$\\
188.6865	&	1724	&$	22.62	\pm	0.03	$&$	22.34	\pm	0.03	$&$	22.43	\pm	0.06	$&$	22.12	\pm	0.09	$\\
344.9195	&	3521	&$	23.28	\pm	0.20	$&$	23.21	\pm	0.15	$&$	23.12	\pm	0.15	$&$	23.29	\pm	0.17	$\\
438.7789	&	3520	&$	>23.83		$&$	23.80	\pm	0.23	$&$	23.64	\pm	0.28	$&$	>23.48		$\\
872.9788	&	7136	&$	>24.19		$&$	>23.92		$&$	>23.87		$&$	>23.85		$\\
1478.9968	&	7096	&$	>24.04		$&$	>24.24		$&$	>23.88		$&$	>24.10		$\\
4832.0789	&	7182	&$	>24.81		$&$	>25.50		$&$	>24.02		$&$	>24.37		$\\
\hline
\end{tabular}
\begin{list}{}{}
\item[$^{\mathrm{(a)}}$] Corrected for Galactic foreground reddening.
\end{list}
\end{table*}

\begin{table*}
\caption{GROND $JHK_s$ photometric data}             
\label{091029JHK}      
\centering                         
\begin{tabular}{r c c c c}        
\hline\hline   
\\            
$T_\mathrm{mid} - T_0$ [ks] & Exposure [s] & \multicolumn{3}{c}{
Brightness$^{\mathrm{(a)}}$ mag$_\mathrm{AB}^{\mathrm{(b)}}$}  \\
 & & $J$ & $H$ & $K_s$  \\
\hline 
0.3135	&	82	&$	17.33	\pm	0.08	$&$	17.26	\pm	0.06	$&$	17.17	\pm	0.11	$\\
0.4171	&	82	&$	17.22	\pm	0.08	$&$	17.13	\pm	0.06	$&$	16.91	\pm	0.09	$\\
0.5221	&	82	&$	17.34	\pm	0.07	$&$	17.23	\pm	0.06	$&$	17.13	\pm	0.11	$\\
0.8835	&	377	&$	17.72	\pm	0.03	$&$	17.61	\pm	0.05	$&$	17.28	\pm	0.07	$\\
1.4625	&	729	&$	18.09	\pm	0.03	$&$	17.94	\pm	0.04	$&$	17.59	\pm	0.08	$\\
2.2379	&	754	&$	18.35	\pm	0.03	$&$	18.45	\pm	0.06	$&$	17.98	\pm	0.10	$\\
3.5383	&	1780	&$	18.63	\pm	0.03	$&$	18.57	\pm	0.05	$&$	18.55	\pm	0.11	$\\
5.3549	&	1772	&$	18.91	\pm	0.04	$&$	18.80	\pm	0.05	$&$	18.68	\pm	0.13	$\\
7.1614	&	1758	&$	18.98	\pm	0.04	$&$	18.86	\pm	0.06	$&$	18.68	\pm	0.11	$\\
8.9684	&	1775	&$	19.01	\pm	0.04	$&$	19.12	\pm	0.06	$&$	18.91	\pm	0.14	$\\
16.0777	&	1777	&$	19.31	\pm	0.05	$&$	19.24	\pm	0.08	$&$	18.98	\pm	0.17	$\\
18.0561	&	1775	&$	19.33	\pm	0.05	$&$	19.30	\pm	0.07	$&$	19.11	\pm	0.19	$\\
104.0267	&	739	&$	21.33	\pm	0.40	$&$	>20.16			$&$	>19.53			$\\
105.2524	&	1644	&$	21.27	\pm	0.29	$&$	>20.48			$&$	>19.70			$\\
170.5336	&	1762	&$	21.79	\pm	0.38	$&$	>20.83			$&$	>19.98			$\\
188.7100	&	1773	&$	21.90	\pm	0.32	$&$	>21.12			$&$	>20.23			$\\
344.9423	&	3569	&$	>21.77			$&$	>21.40			$&$	>20.66			$\\
438.8024	&	3569	&$	>21.68			$&$	>21.10			$&$	>20.80			$\\
873.0025	&	7184	&$	>22.03			$&$	>21.43			$&$	>21.14			$\\
1479.0224	&	7142	&$	>22.07			$&$	>21.51			$&$	>20.98			$\\
4832.1018	&	7230	&$	>22.12			$&$	>21.54			$&$	-			$\\
\hline
\end{tabular}
\begin{list}{}{}
\item[$^{\mathrm{(a)}}$] Corrected for Galactic foreground reddening. Converted
to AB magnitudes for consistency with Table \ref{091029griz}.
\item[$^{\mathrm{(a)}}$] For the SED fitting, the additional error of the absolute
calibration of 0.05 ($J$ and $H$) and 0.07 ($K_s$) mag was added.
\end{list}
\end{table*}

\begin{table*}
\caption{BOOTES-3/YA photometric data}             
\label{091029bootes}      
\centering                         
\begin{tabular}{r c c}        
\hline\hline    
\\            
$T_\mathrm{mid} - T_0$ [ks] & Exposure [s] & Brightness$^{\mathrm{(a)}}$ mag$_\mathrm{AB}$ \\
\hline   
24.8866	&	1440	&$	20.02	\pm	0.11	$\\
27.4346	&	1980	&$	20.09	\pm	0.15	$\\
30.1346	&	1740	&$	20.02	\pm	0.12	$\\
31.9248	&	1800	&$	20.10	\pm	0.10	$\\
36.4357	&	4320	&$	20.22	\pm	0.10	$\\
42.6773	&	5220	&$	20.46	\pm	0.12	$\\
\hline
\end{tabular}
\begin{list}{}{}
\item[$^{\mathrm{(a)}}$] Corrected for Galactic foreground reddening.
\end{list}
\end{table*}

\begin{table*}
\caption{Stardome photometric data}             
\label{091029stardome}      
\centering                         
\begin{tabular}{r c c}        
\hline\hline    
\\            
$T_\mathrm{mid} - T_0$ [ks] & Exposure [s] & Brightness$^{\mathrm{(a)}}$ mag$_\mathrm{AB}$ \\
\hline   
14.2620	&	1800	&$	19.42	\pm	0.20	$\\
16.1568	&	1800	&$	19.56	\pm	0.14	$\\
18.0697	&	1800	&$	19.48	\pm	0.14	$\\
19.9627	&	1800	&$	19.51	\pm	0.16	$\\
21.8566	&	1800	&$	19.82	\pm	0.16	$\\
23.4610	&	1800	&$	19.70	\pm	0.14	$\\
25.6228	&	1800	&$	20.10	\pm	0.14	$\\
30.0153	&	1800	&$	20.13	\pm	0.16	$\\
31.9317	&	1800	&$	20.11	\pm	0.14	$\\
34.4200	&	1800	&$	20.31	\pm	0.12	$\\
37.3110	&	1800	&$	20.26	\pm	0.14	$\\
39.2377	&	1800	&$	20.35	\pm	0.16	$\\
41.1670	&	1800	&$	20.69	\pm	0.15	$\\
43.0920	&	1800	&$	20.64	\pm	0.13	$\\
\hline
\end{tabular}
\begin{list}{}{}
\item[$^{\mathrm{(a)}}$] Corrected for Galactic foreground reddening.
\end{list}
\end{table*}

\end{document}